\documentstyle[manuscript,osa]{revtex}

\begin{document}
\title{Fluctuating Filaments I: Statistical Mechanics of Helices}
\author{S. Panyukov\cite{serg} and Y. Rabin\cite{yit}}
\address{Department of Physics, Bar--Ilan University, Ramat--Gan 52900, Israel }
\maketitle

\begin{abstract}
We examine the effects of thermal fluctuations on thin elastic filaments
with non--circular cross--section and arbitrary spontaneous curvature and
torsion. Analytical expressions for orientational correlation functions and
for the persistence length of helices are derived, and it is found that this
length varies non--monotonically with the strength of thermal fluctuations.
In the weak fluctuation regime, the local helical structure is preserved and
the statistical properties are dominated by long wavelength bending and
torsion modes. As the amplitude of fluctuations is increased, the helix
``melts'' and all memory of intrinsic helical structure is lost. Spontaneous
twist of the cross--section leads to resonant dependence of the persistence
length on the twist rate.
\end{abstract}

\section{Introduction\label{I}}

Modern polymer physics is based on the notion that while real polymers can
be arbitrarily complicated objects, their universal features are captured by
a minimal model in which polymers are described as continuous random walks.
While this approach has been enormously successful and led to numerous
triumphs such as the understanding of rubber elasticity\cite{Flory1}, the
solution of the excluded volume problem and the theory of semi--dilute
polymer solutions\cite{PG79}, it is ill--suited for the description of
non--universal features of polymers which may depend on their chemical
structure in a way that can not be captured by a simple redefinition of the
effective monomer size or its second virial coefficient. For relatively
simple synthetic polymers, such ``local details'' can be treated by polymer
chemistry --type models (e.g., rotational isomer state model\cite{Flory2}).
However, chemically--detailed approaches become prohibitively difficult (at
least as far as analytical modeling is concerned) in the case of complex
biomolecules such as DNA, proteins and their assemblies and a new type of
minimal models is needed to model recent mechanical experiments on such
systems\cite{SFB92,PQSC94,SAB96,HMCL97,Bust,Gaub,MKJ95,Elb,Sack98}. Such an
alternative approach is to model polymers in the way one usually thinks of
them, i.e., as continuous elastic strings or filaments which can be
arbitrarily deformed and twisted. However, while the theory of elasticity of
such objects is well--developed\cite{Love}, little is known about the
statistical mechanics of fluctuating filaments with arbitrary natural
shapes. The main difficulty is mathematical in origin: the description of
three--dimensional filaments with non--circular cross--section and
non--vanishing spontaneous curvature and twist\cite{GF00}, involves rather
complicated differential geometry\cite{Shi94} and most DNA--related
theoretical studies of such models assumed circular cross--sections and
focused on fluctuations around the straight rod configuration \cite
{MS94,KLNH97,BM98,M98,MN98}.

Recently, we reported a study of the effect of thermal fluctuations on the
statistical properties of filaments with arbitrary spontaneous curvature and
twist\cite{PaYr1}. In this work we present a detailed exposition of the
theory and of its application to helical filaments. In Section \ref{II} we
introduce the description of the spatial configuration of the filament in
terms of a triad of unit vectors oriented along the principal axes of the
filament, and show that all the information about this configuration can be
obtained from the knowledge of a set of generalized torsions. The elastic
energy cost associated with any instantaneous configuration of the filament,
is expressed in terms of the deviations of the generalized torsions that
describe this configuration, from their spontaneous values in some given
stress--free reference state. We use this energy to construct the
statistical weights of the different configurations and show that the
deviations of the generalized torsions behave as Gaussian random noises,
whose amplitudes are inversely proportional to the bare persistence lengths
that characterize the rigidity associated with the different deformation
modes. We then derive the differential equations for the orientational
correlation functions that can be expressed as averages of a rotation matrix
which generates the rotation of the triad vectors as one moves along the
contour of the filament. An expression for the persistence length in terms
of one of the correlators is derived. In Section \ref{III} we apply the
general formalism to helical filaments and derive exact expressions for the
correlators (see Appendix A) and for the effective persistence length of an
untwisted helix. We show that the persistence length is, in general a
non--monotonic function of the amplitudes of thermal fluctuations. We also
show that in the weak fluctuation regime, our exact expressions for the
correlators can be derived from a simplified long--wavelength description of
the helix, which is equivalent to the incompressible rod--like chain model 
\cite{BM98}, and that the fluctuation spectrum is dominated by the Goldstone
modes of this rod--like chain. Analytical expressions for the persistence
length of a spontaneously twisted helix are derived (see Appendix B) and it
is found that this length exhibits resonant--like dependence on the rate of
twist. Finally, in Section \ref{IV} we discuss our results and outline
directions for future research.

\section{General Theory of Fluctuating Filaments\label{II}}

A filament of small but finite and, in general, non--circular
cross--section, is modeled as an inextensible but deformable physical curve
parametrized by a contour length $s$ ($0\leq s\leq L$ where $L$ is the
length of the filament). To each point $s$ one attaches a triad of unit
vectors $\left\{ {\bf t}(s)\right\} $ whose component ${\bf t}_{3}$ is the
tangent vector to the curve at $s$, and the vectors ${\bf t}_{1}{\bf (}s{\bf %
)}\ $and ${\bf t}_{2}{\bf (}s{\bf )}$ are directed along the two axes of
symmetry of the cross--section. The vectors $\left\{ {\bf t}(s)\right\} $,
together with the inextensibility condition $d{\bf x/}ds={\bf t}_{3}$, give
a complete description of the space curve ${\bf x}(s)$, as well as of the
rotation of the cross--section (i.e., twist) about this curve.

The rotation of all the vectors ${\bf t}_{i}$ of the triad as one moves from
point $s$ to point $s^{\prime }$ along the line, is generated by the
rotation matrix ${\bf R(}s,s^{\prime }{\bf )}$%
\begin{equation}
{\bf t}_{i}(s)=\sum_{j}R_{ij}{\bf (}s,s^{\prime }{\bf )t}_{j}(s^{\prime })
\label{R1}
\end{equation}
The rotation matrix has the property 
\begin{equation}
{\bf R(}s,s^{\prime }{\bf )=R(}s,s^{\prime \prime }{\bf )R(}s^{\prime \prime
},s^{\prime }{\bf )}  \label{R01}
\end{equation}
where $s^{\prime \prime }$ is an arbitrary point on the contour of the
filament. It satisfies the equation 
\begin{equation}
\frac{\partial R_{ij}{\bf (}s,s^{\prime }{\bf )}}{\partial s}%
=-\sum_{k}\Omega _{ik}(s)R_{kj}{\bf (}s,s^{\prime }{\bf )}  \label{R2}
\end{equation}
where 
\begin{equation}
\Omega _{ij}=\sum_{k}\varepsilon _{ijk}\omega _{k}  \label{LCT}
\end{equation}
$\varepsilon _{ijk}$ is the antisymmetric tensor and $\left\{ \omega
_{k}\right\} $ will be referred to as generalized torsions, for lack of a
better term. The above equations are supplemented by the ``initial''
condition $R_{ij}{\bf (}s,s{\bf )=}\delta _{ij}$, where $\delta _{ij}$ is
the Kronecker delta function. The formal solution of Eq. (\ref{R2}) is given
by the ``time--ordered'' exponential 
\begin{equation}
{\bf R(}s,s^{\prime }{\bf )=T}_{s}\exp \left( -\int_{s^{\prime
}}^{s}ds^{\prime \prime }{\bf \Omega }(s^{\prime \prime })\right)
=\lim_{\Delta s\rightarrow 0^{+}}e^{-{\bf \Omega }(s_{n})\Delta s}\cdots e^{-%
{\bf \Omega }(s_{2})\Delta s}e^{-{\bf \Omega }(s_{1})\Delta s}  \label{Rto}
\end{equation}
The time--ordering operator with respect to $s,$ ${\bf T}_{s}$ is defined by
the second equality in the above equation, where we broke the interval $%
s-s^{\prime }$ into $n$ parts of length $\Delta s$ each, so that $%
s_{1}=s^{\prime }$ and $s_{n}=s^{\prime }$. The origin of the difficulty in
calculating the above expression is that the matrices ${\bf \Omega (}s{\bf )}
$ and ${\bf \Omega (}s^{\prime }{\bf )}$ do not commute for $s\neq s^{\prime
}$ (this is related to the non--Abelian character of the rotation group in
3d).

Eq. (\ref{R2}) is equivalent to a set of generalized Frenet equations from
which one can calculate the spatial configuration of the filament, given a
set of generalized torsions $\left\{ \omega _{k}\right\} $, 
\begin{equation}
\frac{d{\bf t}_{1}}{ds}=\omega _{2}{\bf t}_{3}-\omega _{3}{\bf t}_{2},\qquad 
\frac{d{\bf t}_{2}}{ds}=-\omega _{1}{\bf t}_{3}+\omega _{3}{\bf t}%
_{1},\qquad \frac{d{\bf t}_{3}}{ds}=\omega _{1}{\bf t}_{2}-\omega _{2}{\bf t}%
_{1}  \label{Fre1}
\end{equation}
Note that in the original Frenet description of space curves in terms of a
unit tangent (which coincides with ${\bf t}_{3}$), normal (${\bf n}$) and
binormal (${\bf b}$), one considers mathematical lines for which it would be
meaningless to define twist about the centerline\cite{Solid}. The Frenet
equations contain only two parameters: the curvature $\kappa $ and torsion $%
\tau $: 
\begin{equation}
\frac{d{\bf b}}{ds}=-\tau {\bf n},\qquad \frac{d{\bf n}}{ds}=-\kappa {\bf t}%
_{3}+\tau {\bf b},\qquad \frac{d{\bf t}_{3}}{ds}=\kappa {\bf n}  \label{Fre2}
\end{equation}
The two frames are related through rotation by an angle $\alpha $ about the
common tangent direction (see Fig. 1), 
\begin{equation}
{\bf t}_{1}={\bf b}\cos \alpha +{\bf n}\sin \alpha ,\qquad {\bf t}_{2}=-{\bf %
b}\sin \alpha +{\bf n}\cos \alpha  \label{t1n}
\end{equation}
Substituting this relation into Eqs. (\ref{Fre1}) and using Eqs. (\ref{Fre2}%
), we relate the generalized torsions $\left\{ \omega _{k}\right\} $ to the
curvature $\kappa $, torsion $\tau $ and twist angle $\alpha $, 
\begin{equation}
\omega _{1}=\kappa \cos \alpha ,\qquad \omega _{2}=\kappa \sin \alpha
,\qquad \omega _{3}=\tau +d\alpha /ds  \label{relfren}
\end{equation}

The theory of elasticity of thin rods\cite{Love} is based on the notion that
there exists a stress--free reference configuration defined by the set of
spontaneous (intrinsic) torsions $\left\{ \omega _{0k}\right\} $. The set $%
\left\{ \omega _{0k}\right\} $ together with Eqs. (\ref{R2}) and (\ref{LCT})
(with $\omega _{k}\rightarrow \omega _{0k}$) completely determines the
equilibrium shape of the filament, in the absence of thermal fluctuations.
Neglecting excluded--volume effects and other non--elastic interactions, it
can be shown \cite{DR00} that the elastic energy associated with some actual
configuration $\left\{ \omega _{k}\right\} $ of the filament is a quadratic
form in the deviations $\delta \omega _{k}=\omega _{k}-\omega _{0k}$%
\begin{equation}
U_{el}\left( \left\{ \delta \omega _{k}\right\} \right) =\frac{kT}{2}%
\int_{0}^{L}ds\sum_{k}a_{k}\delta \omega _{k}^{2}  \label{U1}
\end{equation}
where $T$ is the temperature,$\ k$ is the Boltzmann constant, and $a_{i}$
are bare persistence lengths that depend on the elastic constants and on the
principal moments of inertia with respect to the symmetry axes of the
cross--section, in a model--dependent way. Thus, assuming anisotropic
elasticity (with elastic moduli $E_{i}$) and a particular form of the
deformation, one obtains\cite{DR00} $a_{1}=E_{1}I_{1}/kT$, $%
a_{2}=E_{1}I_{2}/kT$ and $a_{3}=E_{2}(I_{1}+I_{2})/kT$ \ where $I_{i}$ are
the principal moments of inertia. In general, the theory of elasticity of
incompressible isotropic rods with shear modulus $\mu $ yields\cite{Love} $%
a_{1}=3\mu I_{1}/kT$, $a_{2}=3\mu I_{2}/kT$, and $a_{3}=C/kT$ where the
torsional rigidity $C$ is also proportional to $\mu $ and depends on the
geometry of the cross--section\cite{comm} (for an elliptical cross--section
with semi--axes $b_{1}$ and $b_{2}$, $C=\pi \mu
b_{1}^{3}b_{2}^{3}/(b_{1}^{2}+b_{2}^{2})$ ).

The elastic energy $U_{el}\left( \left\{ \delta \omega _{k}\right\} \right) $
determines the statistical weight of the configuration $\left\{ \omega
_{k}\right\} $. The statistical average of any functional of the
configuration $B(\left\{ \omega _{k}\right\} )$ is defined as the functional
integral 
\begin{equation}
\left\langle B\left( \left\{ \omega _{k}\right\} \right) \right\rangle =%
\frac{\int D\left\{ \delta \omega _{k}\right\} B\left( \left\{ \omega
_{k}\right\} \right) e^{-U_{el}\left\{ \delta \omega _{k}\right\} /kT}}{\int
D\left\{ \delta \omega _{k}\right\} e^{-U_{el}\left\{ \delta \omega
_{k}\right\} /kT}}  \label{statav}
\end{equation}
Calculating the corresponding Gaussian path integrals we obtain 
\begin{equation}
\left\langle \delta \omega _{i}(s)\right\rangle =0,\qquad \left\langle
\delta \omega _{i}(s)\delta \omega _{j}(s^{\prime })\right\rangle
=a_{i}^{-1}\delta _{ij}\delta (s-s^{\prime })  \label{cor1}
\end{equation}
We conclude that fluctuations of generalized torsions at two different
points along the filament contour are uncorrelated, and that the amplitude
of fluctuations is inversely proportional to the corresponding bare
persistence length.

The statistical properties of fluctuating filaments are determined by the
orientational correlation functions, which can be expressed as averages of
the elements of the rotation matrix, 
\begin{equation}
\left\langle {\bf t}_{i}(s){\bf t}_{j}(s^{\prime })\right\rangle
=\left\langle R_{ij}{\bf (}s,s^{\prime }{\bf )}\right\rangle
=\sum_{k}\left\langle R_{ik}{\bf (}s,s^{\prime \prime }{\bf )}R_{kj}{\bf (}%
s^{\prime \prime },s^{\prime }{\bf )}\right\rangle  \label{corR}
\end{equation}
The last equality was written using Eq. (\ref{R01}), with $s>s^{\prime
\prime }>s^{\prime }$. Inspection of Eqs. (\ref{Rto}) and (\ref{LCT}), shows
that ${\bf R(}s,s^{\prime \prime })$ depends only the torsions $\omega
_{k}(s_{1})$ with $s>s_{1}>s^{\prime \prime },$ and that ${\bf R(}s^{\prime
\prime },s^{\prime })$ depends only on $\omega _{k}(s_{2})$ with $s^{\prime
\prime }>s_{2}>s^{\prime }.$ Since fluctuations of the torsion in two
non--overlapping intervals are uncorrelated (see Eq. (\ref{cor1})), the
average of the product of rotation matrices splits into the product of their
averages: 
\begin{equation}
\left\langle R_{ij}{\bf (}s,s^{\prime }{\bf )}\right\rangle
=\sum_{k}\left\langle R_{ik}{\bf (}s,s^{\prime \prime }{\bf )}\right\rangle
\left\langle R_{kj}{\bf (}s^{\prime \prime },s^{\prime }{\bf )}\right\rangle
\label{RR}
\end{equation}

In order to derive a differential equation for the averaged rotation matrix,
we consider the limit $\Delta s=s-s^{\prime \prime }\rightarrow 0$. Keeping
terms to first order in $\Delta s$ we find 
\begin{equation}
\frac{\partial \left\langle R_{ij}{\bf (}s,s^{\prime }{\bf )}\right\rangle }{%
\partial s}=-\sum_{k}\Lambda _{ik}(s)\left\langle R_{kj}{\bf (}s,s^{\prime }%
{\bf )}\right\rangle  \label{LR}
\end{equation}
where the matrix ${\bf \Lambda }$ is defined as 
\begin{equation}
\Lambda _{ik}\left( s\right) {\bf =}\lim_{\Delta s\rightarrow 0^{+}}\frac{%
\delta _{ik}-\left\langle R_{ik}{\bf (}s,s-\Delta s{\bf )}\right\rangle }{%
\Delta s}  \label{L2}
\end{equation}
Analogously to Eq. (\ref{Rto}), the formal solution of Eq. (\ref{LR}) can be
written as a time--ordered exponential, 
\begin{equation}
\left\langle {\bf R(}s,s^{\prime }{\bf )}\right\rangle {\bf =T}_{s}\exp
\left( -\int_{s^{\prime }}^{s}ds^{\prime \prime }{\bf \Lambda }(s^{\prime
\prime })\right)  \label{toav}
\end{equation}
In order to calculate the matrix ${\bf \Lambda }$ we expand the exponential
in Eq. (\ref{Rto}) to second order in $\Delta s=s-s^{\prime }$ and use the
property of time--ordering operator 
\begin{equation}
{\bf T}_{s}\int_{s-\Delta s}^{s}ds_{1}\int_{s-\Delta s}^{s}ds_{2}{\bf \Omega 
}(s_{1}){\bf \Omega }(s_{2})=\int_{s-\Delta s}^{s}ds_{1}\left[
\int_{s-\Delta s}^{s_{1}}ds_{2}{\bf \Omega }(s_{1}){\bf \Omega }%
(s_{2})+\int_{s_{1}}^{s}ds_{2}{\bf \Omega }(s_{2}){\bf \Omega }(s_{1})\right]
\label{TO}
\end{equation}
In order to average this equation, we first calculate the average of the
product ${\bf \Omega }(s_{1}){\bf \Omega }(s_{2}),$ using Eqs. (\ref{LCT})
and (\ref{cor1}) 
\begin{equation}
\left\langle {\bf \Omega }(s_{1}){\bf \Omega }(s_{2})\right\rangle
=\left\langle {\bf \Omega }(s_{1})\right\rangle \left\langle {\bf \Omega }%
(s_{2})\right\rangle +{\bf diag}(\gamma _{i})\delta (s_{1}-s_{2})
\label{cor2}
\end{equation}
where ${\bf diag}(\gamma _{i})$ is a diagonal matrix with elements 
\begin{equation}
\gamma _{i}=\sum_{k}\frac{1}{2a_{k}}-\frac{1}{2a_{i}}  \label{gammai}
\end{equation}
Using Eqs. (\ref{TO}) and (\ref{cor2}), and keeping terms up to first order
in $\Delta s$ (upon integration, the contribution of $\left\langle {\bf %
\Omega }(s_{1})\right\rangle \left\langle {\bf \Omega }(s_{2})\right\rangle $
is of order $\left( \Delta s\right) ^{2}$), yields 
\begin{equation}
\Lambda _{ik}=\gamma _{i}\delta _{ik}+\sum_{l}\varepsilon _{ikl}\omega _{0l}%
\text{ }  \label{Lam}
\end{equation}

The elements of the averaged rotation matrix are simply the correlators of
the triad vectors (see Eq. (\ref{corR})). From the knowledge of the above
correlators one can calculate other statistical properties of fluctuating
filaments, the most familiar of which is the persistence length $l_{p}$,
that can be interpreted as an effective statistical segment length of a
coarse--grained model, in which one replaces the filament by a random walk
with the same contour length $L$ and rms end--to--end separation $%
\left\langle r^{2}\right\rangle $: 
\[
l_{p}=\lim_{L\rightarrow \infty }\frac{1}{L}\left\langle r^{2}\right\rangle 
\]
The end--to--end vector is defined as ${\bf r=}\int_{0}^{L}{\bf t}_{3}(s)ds$
\ and thus 
\begin{equation}
l_{p}=\lim_{L\rightarrow \infty }\frac{2}{L}\int_{0}^{L}ds\int_{0}^{s}ds^{%
\prime }\left\langle {\bf t}_{3}(s){\bf t}_{3}(s^{\prime })\right\rangle
\label{l}
\end{equation}
The above equations describe the fluctuations of filaments of arbitrary
shape and elastic properties, and in the following this general formalism is
applied to helical filaments.

\section{Fluctuating Helices\label{III}}

\subsection{Untwisted Helix: Correlation Functions and Persistence Length}

Consider a helical filament without spontaneous twist, such that the
generalized spontaneous torsions $\left\{ \omega _{0k}\right\} $ are
independent of position $s$ along the contour. In order to describe the
stress--free configuration of such a filament, it is convenient to introduce
the conventional Frenet triad which consists of the tangent, normal and
binormal to the space curve spanned by the centerline, supplemented by a
constant angle of twist $\alpha _{0}$ which describes the orientation of the
cross--section in the plane normal to the centerline. According to the
general relation between the two frames, Eq. (\ref{relfren}), $\omega
_{01}=\kappa _{0}\cos \alpha _{0}$, $\omega _{02}=\kappa _{0}\sin \alpha
_{0} $ and $\omega _{03}=\tau _{0}$, where $\kappa _{0}$ and $\tau _{0}$ are
the constant curvature and torsion of the space curve in terms of which the
total spontaneous curvature that defines rate of rotation of the helix about
its long axis, is given by $\omega _{0}=(\kappa _{0}^{2}+\tau
_{0}^{2})^{1/2} $. The corresponding helical pitch is $2\pi \tau _{0}/\omega
_{0}^{2}$ and the radius of the helical turn is $2\pi \kappa _{0}/\omega
_{0}^{2}$. We proceed to calculate the orientational correlation functions.

Since ${\bf \Lambda }$ is a constant matrix, Eq. (\ref{toav}) yields (for $%
s_{1}>s_{2}$) 
\begin{equation}
\left\langle {\bf t}_{i}\left( s_{1}\right) {\bf t}_{j}\left( s_{2}\right)
\right\rangle =\left[ e^{-{\bf \Lambda }\left( s_{1}-s_{2}\right) }\right]
_{ij}  \label{tct}
\end{equation}
In order to calculate the matrix $e^{-{\bf \Lambda }\left(
s_{1}-s_{2}\right) }$ we first find the eigenvalues $\lambda _{i}$ of the
matrix ${\bf \Lambda }$, which are determined by the characteristic
polynomial 
\begin{equation}
\lambda ^{3}-\gamma \lambda ^{2}+\mu \lambda -\nu =0  \label{polyn}
\end{equation}
where we introduced the notations 
\begin{eqnarray}
\gamma &=&\gamma _{1}+\gamma _{2}+\gamma
_{3}=a_{1}^{-1}+a_{2}^{-1}+a_{3}^{-1},  \label{eps} \\
\mu &=&\omega _{0}^{2}+\gamma _{1}\gamma _{2}+\gamma _{2}\gamma _{3}+\gamma
_{1}\gamma _{3},  \label{mu} \\
\nu &=&\kappa _{0}^{2}\left( \gamma _{1}\cos ^{2}\alpha _{0}+\gamma _{2}\sin
^{2}\alpha _{0}\right) +\tau _{0}^{2}\gamma _{3}+\gamma _{1}\gamma
_{2}\gamma _{3}  \label{nu}
\end{eqnarray}
The solution of this cubic equation depends on the sign of the expression 
\begin{equation}
\Delta =27\left( \nu -\nu _{1}\right) ^{2}+4\left( \mu -\gamma ^{2}/3\right)
^{3},\qquad \nu _{1}=\frac{1}{3}\gamma \mu -\frac{2}{27}\gamma ^{3}
\label{Delta}
\end{equation}

For $\Delta <0$ all the roots $\lambda _{i}$ are real. In this parameter
range, fluctuations are strong enough to destroy the helical structure on
all length scales. In the limit of very strong fluctuations when the bare
persistence lengths are much smaller than the radii of curvature $\gamma \gg
\omega _{0},$ we have $\lambda _{i}\rightarrow \gamma _{i}$ and correlation
functions become 
\begin{equation}
\left\langle {\bf t}_{i}\left( s_{1}\right) {\bf t}_{j}\left( s_{2}\right)
\right\rangle =e^{-\gamma _{i}\left( s_{1}-s_{2}\right) }\delta _{ij}
\label{strong}
\end{equation}
with $s_{1}-s_{2}>0$. Eq. (\ref{strong}) shows that although angular
correlations remain on length scales smaller than $1/\lambda _{i}$, they are
identical to those of a persistent rod and do not carry any memory of the
original helix.

In the case $\Delta >0$, there is one real eigenvalue, $\lambda _{1}$, and
two complex ones, $\lambda _{2,3}=\lambda _{R}\pm i\omega $, where 
\begin{eqnarray}
\quad \lambda _{1} &=&\frac{K}{6}-2\frac{\mu -\gamma ^{2}/3}{K}+\frac{\gamma 
}{3},\qquad \lambda _{R}=\frac{\gamma -\lambda _{1}}{2}  \label{lambda} \\
\omega &=&\sqrt{3}\left( \frac{K}{12}+\frac{\mu -\gamma ^{2}/3}{K}\right)
,\qquad K=12^{1/3}\left[ 9\left( \nu -\nu _{1}\right) +\sqrt{3\Delta }\right]
^{1/3}  \label{K}
\end{eqnarray}
It is shown in Appendix A that the diagonal orientational correlation
functions take the form 
\begin{equation}
\left\langle {\bf t}_{i}\left( s_{1}\right) {\bf t}_{i}\left( s_{2}\right)
\right\rangle =\left( 1-c_{i}-c_{i}^{\ast }\right) e^{-\lambda _{1}s}+\left(
c_{i}e^{-i\omega s}+c_{i}^{\ast }e^{i\omega s}\right) e^{-\lambda _{R}s}
\label{tt2}
\end{equation}
where $s=s_{1}-s_{2}>0$. The complex coefficients $c_{i}$ are calculated in
Appendix A.

In the limit of small fluctuations, $\gamma \ll \omega _{0}$, we have 
\begin{equation}
\lambda _{1}=\sum_{i}\left( 1-2c_{i}\right) \gamma _{i},\qquad \lambda
_{R}=\sum_{i}c_{i}\gamma _{i},\qquad 2c_{i}=1-\frac{\omega _{0i}^{2}}{\omega
_{0}^{2}},\qquad \omega ^{2}=\omega _{0}^{2}  \label{small}
\end{equation}
In this limit, it is easy to generalize our results for the diagonal
correlators and write down expressions for all the orientational correlation
functions: 
\begin{equation}
\left\langle {\bf t}_{i}(s_{1}){\bf t}_{j}(s_{2})\right\rangle =\frac{\omega
_{0i}\omega _{0j}}{\omega _{0}^{2}}e^{-\lambda _{1}s}+\left( \delta _{ij}-%
\frac{\omega _{0i}\omega _{0j}}{\omega _{0}^{2}}\right) \cos (\omega
_{0}s)e^{-\lambda _{R}s}-\sum_{k}\varepsilon _{ijk}\frac{\omega _{0k}}{%
\omega _{0}}\sin (\omega _{0}s)e^{-\lambda _{R}s}  \label{weak}
\end{equation}
where $s=s_{1}-s_{2}>0$. As expected, Eq. (\ref{weak}) satisfies the
condition of orthonormality of triad vectors ${\bf t}_{i}(s_{1}){\bf t}%
_{j}(s_{1})=\delta _{ij}$ (this geometric condition must be satisfied for
the instantaneous triad vectors, not only on the average). Note that in the
limit of weak fluctuations the local helical structure is preserved on
contour distances $s<\lambda _{R}^{-1}$ and the period of rotation of the
helix about its axis is given by its spontaneous value, $2\pi \omega
_{0}^{-1}$.

Using Eqs. (\ref{eps})--(\ref{Delta}) it can be shown that when $\Delta
\rightarrow 0$, the total curvature of the helix vanishes as $\omega \sim
\Delta ^{1/2}$. Since $\omega \ $is positive for $\Delta >0$ and vanishes
for $\Delta \leq 0$, in a loose sense it plays the role of an order
parameter associated with helical order, and the point $\Delta =0$ can be
interpreted as the critical point at which a continuous helix to random coil
transition takes place. However, although the dependence of $\omega $ on the
various parameters exhibits surprisingly rich behavior, the investigation of
the transition region is of limited physical significance. The change of the
helical period from $2\pi \omega _{0}^{-1}$ to infinity takes place in the
``overdamped'' regime where this period is larger than the persistence
length ($\omega \leq \gamma $), and local helical structure can no longer be
defined in a statistically significant sense. An approximate but more
physically meaningful criterion for the ``melting'' transition is that a
helix of period $2\pi \omega ^{-1}$ melts when the persistence length
becomes of order of this period.

We now return to Eq. (\ref{l}) for the persistence length. Using the matrix
equation $%
\textstyle\int%
_{0}^{\infty }ds\exp (-{\bf \Lambda }s)={\bf \Lambda }^{-1}$ and taking the
appropriate matrix element we find: 
\begin{equation}
l_{p}=2\frac{\tau _{0}^{2}+\gamma _{1}\gamma _{2}}{\kappa _{0}^{2}\left(
\gamma _{1}\cos ^{2}\alpha _{0}+\gamma _{2}\sin ^{2}\alpha _{0}\right)
+\left( \tau _{0}^{2}+\gamma _{1}\gamma _{2}\right) \gamma _{3}}  \label{l1}
\end{equation}
The above expression diverges in the limit of a rigid helix $\gamma
_{i}\rightarrow 0$ in which fluctuations have a negligible effect on the
helix. Non--monotonic behavior is observed for ``plate--like'' helices, with
large radius to pitch ratio, $\kappa _{0}/\tau _{0}$. When no thermal
fluctuations are present ($\gamma _{i}\rightarrow 0$), the effective
persistence length approaches zero. Weak thermal fluctuations ``inflate''
the helix by releasing stored length (by a mechanism similar to the
stretching of the ``slinky'' toy spring) and increase the persistence
length. Eventually, in the limit of strong fluctuations, the persistence
length vanishes again (as $\gamma _{3}^{-1})$ because of the complete
randomization of the filament. Note that the sensitivity to the (constant)
angle of twist increases with radius to pitch ratio.

In the opposite limit of ``rod--like'' helices $\kappa _{0}\rightarrow 0$,
the effective persistence length approaches $2/\gamma _{3}$ and therefore
depends on $a_{1}$ and $a_{2}$ only and not on $\tau _{0}$ and $a_{3}$ which
describe the twist of the cross--section about the centerline. This agrees
with the expectation that since straight inextensible rods do not have
stored length, their end--to--end distance and persistence length are
determined by random bending and torsion (writhe) fluctuations only and are
independent of twist.

\subsection{Weak Fluctuations: The Rod--Like Chain Model}

From the discussion in the preceding section we expect that in the presence
of weak thermal fluctuations, the filament will maintain its helical
structure locally and that fluctuations will only affect its large scale
conformation by introducing random bending and torsion of the helical axis,
as well as random rotation of the filament about this axis. We now rederive
the expressions for the correlators, Eq. (\ref{weak}), using a different
approach that relates our work to that of previous investigators\cite{BM98}
and, in the process, leads to important insights about the nature of the
long wavelength fluctuations that dominate the spectrum of fluctuations in
this regime.

Note that in the absence of thermal fluctuations, $\gamma _{i}=0$, the triad
vectors ${\bf t}_{i}$ attached to the helix can be expressed in terms of the
space--fixed orthonormal triad $\left\{ {\bf e}\right\} $ of vectors ${\bf e}%
_{i}$, where ${\bf e}_{3}$ is oriented along the long axis of the helix and $%
{\bf e}_{1}$ and ${\bf e}_{2}$ lie in the plane normal to it (Fig. 2). It is
convenient to introduce the Euler angles $\phi _{0}(s)=\omega _{0}s,$ $%
\theta _{0}=\arctan (\kappa _{0}/\tau _{0})$ and $\alpha _{0}$ in terms of
which the relation between the two frames is given by 
\begin{equation}
{\bf t}^{R}(s)={\bf R}_{3}(\alpha _{0}){\bf R}_{2}(-\theta _{0}){\bf R}_{3}%
\left[ \phi _{0}(s)\right] {\bf e,}  \label{e2}
\end{equation}
where the rotation matrix 
\begin{equation}
{\bf R}_{3}\left( \phi _{0}\right) =\left( 
\begin{array}{ccc}
\cos \phi _{0} & \sin \phi _{0} & 0 \\ 
-\sin \phi _{0} & \cos \phi _{0} & 0 \\ 
0 & 0 & 1
\end{array}
\right)
\end{equation}
describes rotation by angle $\phi _{0}(s)$ with respect to the ${\bf e}_{3}$
axis. The matrix 
\begin{equation}
{\bf R}_{2}(-\theta _{0})=\left( 
\begin{array}{ccc}
\cos \theta _{0} & 0 & -\sin \theta _{0} \\ 
0 & 1 & 0 \\ 
\sin \theta _{0} & 0 & \cos \theta _{0}
\end{array}
\right)
\end{equation}
gives the rotation by angle $-\theta _{0}$ with respect to the ${\bf e}%
_{2}^{\prime }$ axis (${\bf e}_{2}^{\prime }=$ ${\bf R}_{3}\left[ \phi
_{0}(s)\right] {\bf e}_{2}$), and ${\bf R}_{3}(\alpha _{0})$ is a rotation
by angle $\alpha _{0}$ about the ${\bf e}_{3}^{\prime }$ axis (${\bf e}%
_{3}^{\prime }={\bf R}_{2}(-\theta _{0}){\bf e}_{3}$). Note that while the
space--fixed ${\bf e}$ was taken as a conventional right--handed triad, we
chose the helix--fixed ${\bf t}$ as a left--handed triad. Although this
choice does not affect our previous results, it does affect the geometric
relation between the two coordinate systems and, for consistency, we
replaced the left--handed ${\bf t}$ by the right--handed one, ${\bf t}^{R}=(-%
{\bf t}_{1},{\bf t}_{2},{\bf t}_{3})$, in Eq. (\ref{e2}).

In the presence of weak thermal fluctuations, the axis of the helix slowly
bends and rotates in space, resulting in rotation of the triad $\left\{ {\bf %
e}\right\} $. Since with each point $s$ on the helix we can associate its
projection 
\begin{equation}
\sigma =\tau _{0}s/\omega _{0}  \label{sigma}
\end{equation}
on the long axis of the helix (see Fig. 2), the rotation of the triad $%
\left\{ {\bf e}\right\} $ as one moves along this axis is given by the
generalized Frenet equations, 
\begin{equation}
\frac{d{\bf e}_{1}}{d\sigma }=\varpi _{2}{\bf e}_{3}-\varpi _{3}{\bf e}%
_{2},\qquad \frac{d{\bf e}_{2}}{d\sigma }=-\varpi _{1}{\bf e}_{3}+\varpi _{3}%
{\bf e}_{1},\qquad \frac{d{\bf e}_{3}}{d\sigma }=\varpi _{1}{\bf e}%
_{2}-\varpi _{2}{\bf e}_{1}  \label{stoch}
\end{equation}
The generalized torsions, $\varpi _{i}\left( s\right) $, are Gaussian random
variables determined by the conditions 
\begin{equation}
\left\langle \varpi _{i}\left( \sigma \right) \right\rangle =0,\qquad
\left\langle \varpi _{i}\left( \sigma \right) \varpi _{j}\left( \sigma
^{\prime }\right) \right\rangle =\bar{a}_{i}^{-1}\delta _{ij}\delta \left(
\sigma -\sigma ^{\prime }\right)  \label{dwdw1}
\end{equation}
where the constants $\bar{a}_{i}$ should be determined by the requirement
that the resulting expressions for the correlators (the averages of the
elements of the rotation matrix) coincide with these in Eq. (\ref{weak}). A
calculation similar to that in the previous section yields the correlators 
\begin{equation}
\left\langle {\bf e}_{i}\left( \sigma \right) {\bf e}_{j}\left( \sigma
^{\prime }\right) \right\rangle =\delta _{ij}\exp \left( -\bar{\gamma}%
_{i}\left| \sigma -\sigma ^{\prime }\right| \right) ,  \label{eiej}
\end{equation}
where, analogously to Eq. (\ref{gammai}), we have 
\begin{equation}
\bar{\gamma}_{i}=\sum_{k}\frac{1}{2\bar{a}_{k}}-\frac{1}{2\bar{a}_{i}}
\label{gami}
\end{equation}
Using Eqs. (\ref{e2}), the correlators of the original triad $\left\{ {\bf t}%
\right\} $ can be expressed in terms of the correlators of the $\left\{ {\bf %
e}\right\} $ triad. Comparing the results with Eq. (\ref{weak}), gives 
\begin{equation}
\bar{a}_{1}^{-1}=\bar{a}_{2}^{-1}=\sum_{i}\gamma _{i}\frac{\omega _{0i}^{2}}{%
\omega _{0}\tau _{0}},\qquad \bar{a}_{3}^{-1}=\sum_{i}\frac{1}{a_{i}}\frac{%
\omega _{0i}^{2}}{\omega _{0}\tau _{0}}  \label{a1a3}
\end{equation}
where the equality $\bar{a}_{1}=\bar{a}_{2}$ is the consequence of symmetry
under rotation in the $({\bf e}_{1},{\bf e}_{2})$ plane.

The correlators (\ref{dwdw1}) can be derived from an effective free energy
which describes the long wavelength fluctuations of the helical filament, on
length scales larger than the period of the helix $\omega _{0}^{-1}$. 
\begin{equation}
U_{el}^{LW}=\frac{kT}{2}\int d\sigma \left[ \bar{a}_{1}\left( \varpi
_{1}^{2}+\varpi _{2}^{2}\right) +\bar{a}_{3}\varpi _{3}^{2}\right]
\label{RLC}
\end{equation}
This expression coincides with the elastic energy of a rod--like chain (RLC)
introduced by Bouchiat and Mezard\cite{BM98}. The persistence length $\bar{a}%
_{1}$ describes the elastic response to bending and torsion of the effective
rod--like filament. The persistence length $\bar{a}_{3}$ controls the
elastic response of the RLC to twist about its axis. As a consequence of the
fluctuation--dissipation theorem, it also determines the amplitude of
fluctuations $\Delta \phi $ of the angle $\phi (\sigma )=\omega
_{0}^{2}\sigma /\tau _{0}+\Delta \phi (\sigma )$, where the correlator of
the random angle of rotation about the axis of the RLC is given by 
\begin{equation}
\left\langle \left[ \Delta \phi (\sigma )-\Delta \phi (\sigma ^{\prime })%
\right] ^{2}\right\rangle =\bar{a}_{3}^{-1}\left| \sigma -\sigma ^{\prime
}\right|  \label{randang}
\end{equation}

In Eq. (\ref{a1a3}) we calculated the effective persistence lengths of this
model ($\bar{a}_{i}$) in terms of the bare parameters of the underlying
helical filament. In reference \cite{BM98} where the analysis begins with
the RLC model, these corresponding persistence lengths were introduced by
hand. The difference between the two models becomes important if one
considers the combined application of extension and twist: while such a
coupling appears trivially in models of stretched helical filaments\cite
{DR00}, twist has no effect on the extension in the RLC model\cite{BM98}, in
contradiction with experimental observations \cite{SAB96}. Our analysis
underscores the fact that the RLC model does not give a complete description
of the fluctuating helix. Rather, it describes long wavelength fluctuations
of the ``phantom'' axes $\left\{ {\bf e}_{i}\right\} $ which, by themselves,
contain no information about the local helical structure of the filament. In
order to recover this information and construct the correlators of the
original helix $\left\langle {\bf t}_{i}\left( s_{1}\right) {\bf t}%
_{i}\left( s_{2}\right) \right\rangle $, one has to go beyond the RLC model
and reconstruct the local helical geometry using the relation between ${\bf e%
}_{i}$ and the helix--fixed axes ${\bf t}_{i}$, Eqs. (\ref{e2}).

In deriving the expressions for the correlators $\left\langle {\bf t}_{i}(s)%
{\bf t}_{j}(0)\right\rangle $ in terms of the correlators of the RLC model,
we did not take into account the possibility of fluctuations of the twist
angle of the cross--section of the helix about its centerline, $\alpha
_{0}\rightarrow \alpha (s)=\alpha _{0}+\Delta \alpha (s)$. From the fact
that the resulting correlators coincide with the exact expressions, Eq. (\ref
{weak}), we conclude that such fluctuations do not contribute to the
correlators. This surprising result follows from the fact that in the weak
fluctuation regime, the statistical properties of the helix are completely
determined by the low energy part of the fluctuation spectrum. Such
long--wavelength fluctuation modes (Goldstone modes) lead to the loss of
helical correlations on length scales larger than all the natural length
scales of the helix ($s\geq \gamma ^{-1}\gg \omega _{0}^{-1}$). These
Goldstone modes are associated with spontaneously broken continuous
symmetries and correspond to bending ($\varpi _{1}$ and $\varpi _{2}$) and
twist ($\varpi _{3}$) modes of the RLC. It is important to emphasize that
these modes correspond to different deformations of the centerline of the
helix and not to twist of its cross--section about this centerline. Since
the elastic energy, Eq. (\ref{U1}), depends on the spontaneous angle of
twist of the helix about its centerline through the combinations $\delta
\omega _{1}=\kappa \cos \alpha -\kappa _{0}\cos \alpha _{0}$ and $\delta
\omega _{2}=\kappa \sin \alpha -\kappa _{0}\sin \alpha _{0},$ we conclude
that the energy is not invariant under global rotation of the cross--section
about the centerline and that such a rotation is not a continuous symmetry
of the helix. Therefore, twist fluctuations of the helical cross--section
are not Goldstone modes and do not contribute to the correlators in the weak
fluctuation limit.

Another interesting observation is that there is no contribution from
compressional modes to the long--wavelength energy, Eq. (\ref{U1}). This is
surprising since the RLC is a coarse--grained representation of the helix
and the latter may be expected to behave as a compressible object, with
accordion--like compressional modes\cite{M98}. In order to check this point,
we write down the spatial position of a point $s$ on the helix as 
\begin{equation}
{\bf x}(s)={\bf \bar{x}}(\sigma )+\delta {\bf x}(s)  \label{xsig}
\end{equation}
where ${\bf \bar{x}}(\sigma )$ describes the curve spanned by the long axis
of the helix and, therefore, defines the spatial position of the point $%
\sigma $, Eq. (\ref{sigma}), on the RLC contour. The deviation $\delta {\bf x%
}(s)$ describes the rotation of the locally helical filament about this
axis. Since the original filament is incompressible, it satisfies $d{\bf x/}%
ds={\bf t}_{3}$. From Eq. (\ref{e2}) we obtain an expression for ${\bf t}%
_{3} $ which, upon substitution into the incompressibility condition and
averaging over length scales $\left\{ |\varpi _{i}|^{-1}\right\} \gg s$ $\gg
\omega _{0}^{-1}$ (much larger than the inverse total curvature of the helix
but much smaller than the radii of curvature of the RLC), yields 
\begin{equation}
\frac{d{\bf \bar{x}}(\sigma )}{d\sigma }={\bf e}_{3}(\sigma )
\label{compeff}
\end{equation}
The fact that the long--wavelength fluctuations of the helix satisfy the
above incompressibility conditions, implies that compressional fluctuations
do not contribute to the long--wavelength correlators. The origin of this
observation becomes clear if we recall that the energy of the helix depends
on the spontaneous curvature $\kappa _{0}$ and torsion $\tau _{0}$ and,
since compressional modes change the local curvature and torsion, they have
a gap in the energy spectrum and their energy does not vanish even in the
long--wavelength limit. We conclude that similarly to twist fluctuations of
the helical cross--section, compressional modes are not Goldstone modes.

The above deliberations have profound consequences for the elastic response
of the filament to long--wavelength perturbations, such as tensile forces
and moments applied to its ends. Using the fluctuation--dissipation theorem,
we conclude that as long as the deformation of the filament remains small
(on scale $\omega _{0}^{-1}$), these forces and moments do not induce twist
of the cross--section of the helix about its centerline, and that the
deformation can be completely described by the incompressible RLC model.

\subsection{Effect of Spontaneous Twist}

We proceed to calculate the persistence length of a helix whose
cross--section is twisted by an angle $\alpha _{0}(s)=\dot{\alpha}_{0}s$
about the centerline ($\dot{\alpha}_{0}$ is a constant rate of twist). It is
convenient to rewrite Eq. (\ref{l}) as: 
\begin{equation}
l_{p}=\lim_{L\rightarrow \infty }\frac{2}{L}\int_{0}^{L}ds^{\prime
}\int_{0}^{L-s^{\prime }}ds\left\langle {\bf t}_{3}(s+s^{\prime }){\bf t}%
_{3}(s^{\prime })\right\rangle  \label{l2}
\end{equation}
Recall that the correlator in the integrand of Eq. (\ref{l2}) is simply the $%
33$ element of the averaged rotation matrix, and is therefore the solution
of equation Eq. (\ref{LR}), the coefficients of which are the elements of
the matrix ${\bf \Lambda }(s+s^{\prime })$ defined in Eq. (\ref{Lam}). The
diagonal elements of this matrix are constants ($\gamma _{i}$), while the
non--diagonal elements are given by the expressions 
\begin{eqnarray}
\Lambda _{12}(s+s^{\prime }) &=&-\Lambda _{21}(s+s^{\prime })=\tau _{0}+\dot{%
\alpha}_{0},\qquad  \nonumber \\
\Lambda _{31}(s+s^{\prime }) &=&-\Lambda _{13}(s+s^{\prime })=\kappa
_{0}\sin \left( \dot{\alpha}_{0}s+\alpha _{0}\right) ,  \nonumber \\
\text{\ }\Lambda _{23}(s+s^{\prime }) &=&-\Lambda _{32}(s+s^{\prime
})=\kappa _{0}\cos \left( \dot{\alpha}_{0}s+\alpha _{0}\right) ,
\label{last}
\end{eqnarray}
where all the dependence on $s^{\prime }$ is contained in $\alpha
_{0}=\alpha _{0}(s^{\prime }).$

The correlator in Eq. (\ref{l2}) decays exponentially fast with $s$, and
thus the upper limit on the integral over $s$ can be extended to infinity.
Since the correlator is a periodic function of $\alpha _{0},$ the
integration over $s^{\prime }$ can be replaced by that over $\alpha _{0}$
and we obtain 
\begin{equation}
l_{p}=\int_{0}^{2\pi }\frac{d\alpha _{0}}{\pi }\int_{0}^{\infty
}ds\left\langle {\bf t}_{3}(s){\bf t}_{3}(s-s_{1})\right\rangle
\label{persist}
\end{equation}
In deriving the above expression we assumed that the limit $L\rightarrow
\infty $ is taken and that the total angle of twist is always large, $L\dot{%
\alpha}_{0}$ $\gg 2\pi $ (i.e., the product $L\dot{\alpha}_{0}$ remains
finite for arbitrarily small $\dot{\alpha}_{0}$). This assumption will be
used in the following analysis.

We first consider some limiting cases in which analytical results can be
derived. In the limit of vanishing twist rates, $\dot{\alpha}_{0}\rightarrow
0,$ the persistence length is obtained by averaging Eq. (\ref{l1}) with
respect to $\alpha _{0}$. This yields: 
\begin{equation}
l_{p}=%
{\displaystyle{2(\tau _{0}^{2}+\gamma _{+}^{2}-\gamma _{-}^{2}) \over \sqrt{[\kappa _{0}^{2}\gamma _{+}+(\tau _{0}^{2}+\gamma _{+}^{2}-\gamma _{-}^{2})\gamma _{3}]^{2}-\kappa _{0}^{4}\gamma _{-}^{2}}}}%
\label{lp3}
\end{equation}
where 
\begin{equation}
\gamma _{\pm }\equiv (\gamma _{1}\pm \gamma _{2})/2  \label{gammp}
\end{equation}
with $\gamma _{1\text{ }}$and $\gamma _{2}$ defined in (\ref{gammai}).

In the limit of large twist rates, $\dot{\alpha}_{0}\rightarrow \infty ,$ we
can replace the denominator of Eq. (\ref{l1}) by its average with respect to 
$\alpha _{0}$. This yields 
\begin{equation}
l_{p}=\frac{2\left( \tau _{0}^{2}+\gamma _{+}^{2}-\gamma _{-}^{2}\right) }{%
\kappa _{0}^{2}\gamma _{+}+\left( \tau _{0}^{2}+\gamma _{+}^{2}-\gamma
_{-}^{2}\right) \gamma _{3}}  \label{lp4}
\end{equation}
Finally, when $\gamma _{1}=\gamma _{2}$ ($a_{1}=a_{2}$), the persistence
length becomes independent of twist and can be derived from either of Eqs. (%
\ref{lp3}) and (\ref{lp4}), by substituting $\gamma _{-}=0.$

We now consider the case of arbitrary twist rates and fluctuation
amplitudes. The calculation involves the solution of linear differential
equations with periodic coefficients and details are given in Appendix B. We
obtain: 
\begin{equation}
l_{p}=\frac{2\gamma _{3}^{-1}}{1+\left( \Xi -1\right) ^{-1}+\left( \Xi
^{\ast }-1\right) ^{-1}}  \label{plth}
\end{equation}
An analytical expression for the complex function $\Xi (\dot{\alpha}_{0})$
is given in Appendix B.

In Fig. 3 we present a three--dimensional plot of the persistence length
given in units of the helical pitch $l^{\ast }=l\omega _{0}^{2}/2\pi \tau
_{0},$ as a function of the dimensionless rate of twist $w=2\omega _{0}^{-1}%
\dot{\alpha}_{0}$ and of the logarithm of the bare persistence length $a_{1}$%
, for a ``plate--like'' helix with large radius to pitch ratio $\kappa
_{0}/\tau _{0}$. Inspection of Fig. 3 shows that in the case of a circular
cross--section with $a_{1}=a_{2}=1000$, the persistence length becomes
independent of twist. With increasing asymmetry, $a_{1}<a_{2}$, a maximum
appears at vanishing twist rates, accompanied by two minima at $\dot{\alpha}%
_{0}=\pm \omega _{0}/2$. The geometrical significance of the locations ($%
\dot{\alpha}_{0}=0,\pm \omega _{0}/2$) of these resonances is underscored by
the observation that in the limit of vanishing pitch, a ribbon--like
untwisted ($\dot{\alpha}_{0}=0$) helix degenerates into a ring. For $\dot{%
\alpha}_{0}=\pm \omega _{0}/2$, the cross--section of a twisted helix
rotates by $\pm \pi $ with each period, and in the above limit the helix
degenerates into a M\H{o}bius ring. As asymmetry increases ($a_{1}\ll a_{2}$%
), each extremum splits into a minimum and a maximum and eventually one
obtains a dip at $\dot{\alpha}_{0}=0$, accompanied by two symmetrical peaks
at $\dot{\alpha}_{0}\simeq \pm \omega _{0}/2$. Note that the persistence
length is a non--monotonic function of the amplitude of thermal fluctuations
(i.e., of $1/a_{1}$): it first slowly increases and eventually decreases
rapidly with decreasing $a_{1}.$ Several two--dimensional plots of the
persistence length as a function of the rate of twist, for different
combinations of the bare persistence lengths $a_{i}$ are shown in Fig. 4.
The detailed behavior of the persistence length depends sensitively on the
choice of the parameters: for example, in the limit of weak fluctuations
three maxima are observed in Fig. 4, instead of a maximum accompanied by two
minima in Fig. 3. In all cases, the locations of the extrema are determined
by geometry only: $\dot{\alpha}_{0}=0$, $\pm \omega _{0}/2$.

In order to demonstrate how the initial choice of the handedness of the
helix breaks the symmetry between the effects of under and over--twist on
the persistence length, in Fig. 5 we present a three--dimensional plot of
the persistence length as a function of the dimensionless rate of twist $w$
and of the inverse radius of curvature $\kappa _{0}$, for helices with
radius to pitch ratios of order unity and large asymmetry of the
cross--section, $a_{1}\gg a_{2}$. Note that for $\kappa _{0}/\tau _{0}<1$
(rod--like helices), there is a single broad maximum at $\dot{\alpha}%
_{0}=-\omega _{0}/2$. Then, at $\kappa _{0}/\tau _{0}\simeq 1,$ a central
peak appears at $\dot{\alpha}_{0}=0$. This peak grows much faster than the $%
\dot{\alpha}_{0}=-\omega _{0}/2$ peak, with increasing $\kappa _{0}/\tau
_{0}.$ At yet higher values of $\kappa _{0}/\tau _{0}$ another peak appears
at $\dot{\alpha}_{0}$ $=\omega _{0}/2$ and eventually the amplitudes of the
two M\H{o}bius side--peaks become equal (and much smaller than the amplitude
of the $\dot{\alpha}_{0}$=$0$ peak) in the limit of plate--like helices, $%
\kappa _{0}/\tau _{0}\gg 1$ (see curve 1 in Fig. 4).

What is the origin of the M\H{o}bius resonances observed in Figs. 3--5?
Recall that the calculation of the persistence length of a twisted helix
involves the solution of linear differential equations with periodic
coefficients (Eqs. (\ref{dgi}) in Appendix B). These equations were derived
from linear differential equations with periodic coefficients and
multiplicative random noise, Eqs. (\ref{R2}) and Eqs. (\ref{Fre1}), which
are known to lead to stochastic resonances\cite{Git99}. Some physical
intuition can be derived from the following argument. While the persistence
length is a property of the space curve described by the Frenet triad, the
microscopic Brownian motion of the filament arises as the result of random
forces that act on its cross--section and therefore are given in the frame
associated with the principal axes of the filament. Since the two frames are
related by a rotation of the cross--section by an angle $\alpha _{0}(s)$,
the random force in the Frenet frame is modulated by linear combinations of $%
\sin \alpha _{0}(s)$ and $\cos \alpha _{0}(s)$. This gives a deterministic
contribution to the persistence length which, to lowest order in the force,
is proportional to the mean square amplitude of the random force and
therefore varies sinusoidally with $\pm 2\alpha _{0}(s)$. The M\H{o}bius
resonances occur whenever the total curvature of the helix $\omega _{0}$
coincides with the rate of variation of this deterministic contribution of
the random force, $\pm 2\dot{\alpha}_{0}$.

\section{Discussion\label{IV}}

In this work we studied the statistical mechanics of thermally fluctuating
elastic filaments with arbitrary spontaneous curvature and twist. We
constructed the equations for the orientational correlation functions and
for the persistence length of such filaments. We would like to stress that
our theory describes arbitrarily large deviations of a long filament from
its equilibrium shape; the only limitation is that fluctuations are small on
microscopic length scales, of the order of the thickness of the filament.
Furthermore, since the equilibrium shape and the fluctuations of the
filaments are completely described by the set of spontaneous torsions $%
\left\{ \omega _{0k}\right\} $ and its fluctuations $\left\{ \delta \omega
_{k}\right\} $ respectively, our theory is set up in the language of
intrinsic geometry of the space curves. All the interesting statistical
information is contained in the correlators of the triad vectors $\left\{ 
{\bf t}\right\} $ which can be expressed in terms of the known correlators
of the fluctuations $\left\{ \delta \omega _{k}\right\} $, using the Frenet
equations. Since these equations describe pure rotation of the triad
vectors, this has the advantage that fluctuations of the torsions introduce
only random rotations of the vectors of the triad, and preserve their unit
norm. The use of intrinsic geometry automatically ensures that the
inextensibility constraint is not violated in the process of thermal
fluctuations and therefore does not even have to be considered explicitly in
our approach. We would like to remind the readers that the formidable
mathematical difficulties associated with attempts to introduce this
constraint, have hindered the development of persistent chain type models in
the past and led to the introduction of the mean spherical approximation in
which the constraint is enforced only on the average, and to perturbative
expansions about the straight rod limit.

The general formalism was then applied to helical filaments both with and
without twist of the cross--section about the centerline. In the latter case
we found that weak thermal fluctuations are dominated by long wavelength
Goldstone modes that correspond to bending and twist of the coarse--grained
filament (the rod--like chain). Such fluctuations distort the helix on
length scales much larger than its natural period but do not affect its
local structure and, in particular, do not change the angle of twist of the
cross--section about the centerline. Strong thermal fluctuations lead to
melting of the helix, accompanied by complete loss of local helical
structure. Depending on the parameters of the helix, the persistence length
is a non--monotonic function of the strength of thermal fluctuations, and
may first increase and then decrease as the amplitude of fluctuations is
increased. Resonant peaks and dips in plots of the persistence length versus
the spontaneous rate of twist are observed both for small twist rates and
for rates equal to half the total curvature of the helix, phenomena which
bear some formal similarity to stochastic resonances.

There are several possible directions in which the present work can be
extended. We did not consider here the effects of excluded volume and other
non--elastic interactions, on the statistical properties of fluctuating
filaments. Such an analysis requires the introduction of a field theoretical
description of the filaments\cite{PaYr2}. While this approach is interesting
in its own right, we expect that the excluded volume exponent for the
scaling of the end--to--end distance of a single filament will be identical
to that of a Gaussian polymer chain (self--avoiding random walk). However,
new effects related to liquid crystalline ordering are expected in dense
phases of such filaments. Another possible extension of the model relates to
the elasticity of random heteropolymers, with quenched distribution of
elastic constants and/or spontaneous torsions\cite{Nel98}.

A\ natural application of our theory involves the modeling of mechanical
properties and conformational statistics of chiral biomolecules such as DNA
and RNA. The advantage of our theory is that it allows us to take into
account, in an exact manner, the effects of thermal fluctuations on the
persistence length and other elastic parameters of the filament. Thus, the
generalization of the theory to include the effect of tensile forces and
torques applied to the ends of the filament, is expected to lead to new
predictions for mechanical stretching experiments in the intermediate
deformation regime, for tensile forces that affect the global but not the
local (on length scales $\leq $ $l_{p}$) conformation of the filament.
Measurements of the effect of elongation on thermal fluctuations of the
molecule, can give information about its elastic constants, and help resolve
long--standing questions regarding the natural curvature of DNA\cite
{TTH87,BFK95}. It is interesting to compare our expression for the
persistence length to that introduced by Trifonov et al.\cite{TTH87} who
proposed that the apparent persistence length $l_{a}$ of DNA depends not
only on the rigidity (dynamic persistence length $l_{d}$), but also on the
intrinsic curvature of the\ molecule\ (static persistence length $l_{s}$).
The apparent persistence length is given in terms of the two others as 
\begin{equation}
\frac{1}{l_{a}}=\frac{1}{l_{d}}+\frac{1}{l_{s}}  \label{trif}
\end{equation}
Note that the philosophy of the above approach is very similar to ours -- we
begin with filaments which have some given intrinsic length (spontaneous
radius of curvature/torsion), and find that the interplay between this
length and thermal fluctuations gives rise to a persistence length $l_{p}$.
In fact, taking for simplicity the case of a circular cross--section, $%
a_{1}=a_{2}$, our expression Eq. (\ref{l1}), can be recast into the form of
Eq. (\ref{trif}), with 
\begin{equation}
l_{a}=l_{p},\text{\qquad }l_{d}=2a_{1},\qquad l_{s}=\kappa _{0}^{-2}\left(
\gamma _{1}+\tau _{0}^{2}/\gamma _{1}\right)  \label{trif1}
\end{equation}
Indeed, in our model, $a_{1}$ is the bare persistence length that determines
the length scale on which the filament is deformed by thermal bending and
torsion fluctuations. Our analog of the static persistence length $l_{s}$
depends on the spontaneous bending rate $\kappa _{0}$ and diverges in the
case of a straight filament ($\kappa _{0}\rightarrow 0$), in which case $%
l_{a}\rightarrow l_{d}$. If we make the further assumption that twist
rigidity is much smaller than the bending rigidity, $a_{3}\ll a_{1}$, the
static persistence length becomes independent of the bending rigidity and
depends on both the spontaneous curvature and the twist rigidity. Note,
however, that the resulting $\kappa _{0}^{-2}$ dependence of $l_{s}$ differs
from the originally proposed one ($\kappa _{0}^{-1}$)\cite{TTH87}$.$

Another possible application of our theory involves a new way of looking
into the protein folding problem. Usually, one assumes that the folded
conformation of proteins is determined by the interactions between the
constituent amino--acids. A different approach, more closely related to the
present work, would be to reverse the common logic: instead of trying to
understand what kind of spatial structure will result for a given primary
sequence of amino--acids, one can begin with a known equilibrium shape
(native state) and attempt to identify the parameters of an effective
filament (distributions of spontaneous torsions $\left\{ \omega
_{0i}(s)\right\} $ ) which will give rise to this three--dimensional
structure\cite{Shura}. Knowledge about the fluctuations and the melting of
proteins can then be used to determine the distribution of the bare
persistence lengths $\left\{ a_{i}(s)\right\} $. While the question of
whether such an approach can be successfully implemented in order to
determine the relation between primary sequence and ternary structure
remains open, our insights about the statistical properties of fluctuating
filaments are clearly applicable to modeling of $\alpha -$helices and other
elements (e.g., $\beta -$sheets) of secondary structure of proteins.

\medskip

\begin{quotation}
{\Large Appendix A: Calculation of Correlation Functions}\label{A}
\end{quotation}

We begin with the construction of the eigenvectors of the matrix ${\bf %
\Lambda }$, defined by Eq. (\ref{Lam}), in the case $\Delta >0$ (see Eq. (%
\ref{Delta})), when there is one real eigenvalue $\lambda _{1}$ and two
complex ones, $\lambda _{R}\pm i\omega $. Expanding this matrix over its
eigenvectors, we get 
\begin{equation}
\Lambda _{ij}=\lambda _{1}\bar{u}_{i}u_{j}+\left( \lambda _{R}+i\omega
\right) \bar{v}_{i}v_{j}^{\ast }+\left( \lambda _{R}-i\omega \right) \bar{v}%
_{i}^{\ast }v_{j}  \label{A1}
\end{equation}
where the eigenvectors ${\bf u}$, ${\bf \bar{u}}$, ${\bf v}$, ${\bf \bar{v}}$
(and the complex conjugates of the latter two, ${\bf v}^{\ast }$ and ${\bf 
\bar{v}}^{\ast }$) obey the orthonormality conditions 
\begin{equation}
\sum_{i=1}^{3}\bar{u}_{i}u_{i}=\sum_{i=1}^{3}\bar{v}_{i}v_{i}^{\ast
}=1,\qquad \sum_{i=1}^{3}\bar{u}_{i}v_{i}=\sum_{i=1}^{3}\bar{v}%
_{i}u_{i}=\sum_{i=1}^{3}\bar{v}_{i}v_{i}=0  \label{ortho}
\end{equation}
Using these conditions we can exponentiate the matrix ${\bf \Lambda }$%
\begin{equation}
\left[ e^{-{\bf \Lambda }s}\right] _{ij}=\bar{u}_{i}u_{j}e^{-\lambda _{1}s}+%
\bar{v}_{i}v_{j}^{\ast }e^{-\left( \lambda _{R}+i\omega \right) s}+\bar{v}%
_{i}^{\ast }v_{j}e^{-\left( \lambda _{R}-i\omega \right) s}  \label{expL}
\end{equation}

Since we are interested only in the diagonal elements of this matrix, it is
convenient to introduce the notations 
\begin{equation}
c_{i}=\bar{v}_{i}v_{i}^{\ast },\qquad \sum_{i=1}^{3}c_{i}=1  \label{A2}
\end{equation}
In addition, substituting $s=0$ in Eq. (\ref{expL}) we get 
\begin{equation}
\bar{u}_{i}u_{i}=1-c_{i}-c_{i}^{\ast }  \label{A3}
\end{equation}
In order to find the complex coefficients $c_{i}$ we write down expressions
for diagonal elements of the matrices ${\bf \Lambda }$ and ${\bf \Lambda }%
^{2}$%
\begin{equation}
\begin{array}{c}
\gamma _{i}=\left( 1-c_{i}-c_{i}^{\ast }\right) \lambda _{1}+c_{i}\left(
\lambda _{R}+i\omega \right) +c_{i}^{\ast }\left( \lambda _{R}-i\omega
\right) \\ 
\left( \gamma _{i}-\lambda _{1}\right) ^{2}-\omega _{0}^{2}+\omega
_{0i}^{2}=\left( 1-c_{i}-c_{i}^{\ast }\right) \lambda _{1}^{2}+c_{i}\left(
\lambda _{R}+i\omega \right) ^{2}+c_{i}^{\ast }\left( \lambda _{R}-i\omega
\right) ^{2}
\end{array}
\label{A4}
\end{equation}
Looking for the solution of these equations in the form $c_{i}=%
\mathop{\rm Re}%
c_{i}+i%
\mathop{\rm Im}%
c_{i}$ we get expressions for real and imaginary parts of complex parameters 
$c_{i}$%
\begin{equation}
\begin{array}{c}
2%
\mathop{\rm Re}%
c_{i}=%
{\displaystyle{-\gamma _{i}^{2}+2\varepsilon _{i}\left( \lambda _{1}+\lambda _{R}\right) +2\lambda _{R}\allowbreak \lambda _{1}+\omega _{0}^{2}-\omega _{0i}^{2} \over \omega ^{2}+\left( \lambda _{1}-\lambda _{R}\right) ^{2}}}%
, \\ 
2\omega 
\mathop{\rm Im}%
c_{i}=\lambda _{1}-\gamma _{i}+2\left( \lambda _{R}-\lambda _{1}\right) 
\mathop{\rm Re}%
c_{i}
\end{array}
\label{A5}
\end{equation}

{\Large Appendix B: Persistence Length of Twisted Helix}\label{B}

Since the persistence length is defined by the $33$ element of the averaged
rotation matrix, we will consider the $i3$ component of equation Eq. (\ref
{LR}) which, using Eq. (\ref{corR}), can be expressed as an equation for the
corresponding correlator: 
\begin{equation}
\frac{dg_{i}}{ds}=-\sum_{l}\Lambda _{il}(s+s^{\prime })g_{l},\qquad
g_{i}(s,s^{\prime })\equiv \left\langle {\bf t}_{i}(s+s^{\prime }){\bf t}%
_{3}(s^{\prime })\right\rangle  \label{dgi}
\end{equation}
with initial conditions $g_{1}\left( 0,s^{\prime }\right) =g_{2}\left(
0,s^{\prime }\right) =0$ and $g_{3}\left( 0,s^{\prime }\right) =1$. The
matrix ${\bf \Lambda }\left( s+s^{\prime }\right) $ was defined in Eq. (\ref
{last}). Note that since the only $s$--dependent parameter of the helix is
the angle of twist, the correlators $g_{i}(s,s^{\prime })$ depend on $%
s^{\prime }$ only through the parameter $\alpha _{0}(s^{\prime })=\alpha
_{0} $ and, in order to simplify the notation, we will omit the second
argument of these functions in the following.

It is convenient to introduce the complex function 
\begin{equation}
f\left( s\right) =\left[ g_{1}(s)+ig_{2}(s)\right] e^{-i\left( \dot{\alpha}%
_{0}s+\alpha _{0}\right) }  \label{f(s)}
\end{equation}
such that $f$ and $g_{3}$ obey the coupled equations 
\begin{eqnarray}
\frac{df}{ds}+\gamma _{+}f+\gamma _{-}f^{\ast }e^{-2i\left( \dot{\alpha}%
_{0}s+\alpha _{0}\right) } &=&-i\kappa _{0}g_{3}+i\tau _{0}f,  \nonumber \\
\frac{dg_{3}}{ds}+\gamma _{3}g_{3} &=&-i\kappa _{0}%
{\displaystyle{1 \over 2}}%
\left( f-f^{\ast }\right)  \label{fcoup}
\end{eqnarray}
Taking a Laplace transform of these equations, 
\begin{equation}
\tilde{f}\left( p\right) \equiv \int_{0}^{\infty }f\left( s\right)
e^{-ps}ds,\qquad \tilde{g}_{3}\left( p\right) \equiv \int_{0}^{\infty
}g_{3}\left( s\right) e^{-ps}ds  \label{Lapl}
\end{equation}
where $p$ is, in general, a complex parameter, we get 
\begin{eqnarray}
\left( p+\gamma _{+}-i\tau _{0}\right) \tilde{f}\left( p\right) +i\kappa _{0}%
\tilde{g}_{3}\left( p\right) &=&-\gamma _{-}e^{-2i\alpha _{0}}\tilde{f}%
^{\ast }\left( p+2i\dot{\alpha}_{0}\right) ,  \label{Laplf} \\
\left( p+\gamma _{3}\right) \tilde{g}_{3}\left( p\right) +i\kappa _{0}%
{\displaystyle{1 \over 2}}%
\left[ \tilde{f}\left( p\right) -\tilde{f}^{\ast }\left( p\right) \right]
&=&1  \label{Laplg}
\end{eqnarray}
In deriving these equations, we used the initial conditions, $f\left(
0\right) =0$ and $g_{3}\left( 0\right) =1.$ Substituting $\tilde{g}_{3}$
from Eq. (\ref{Laplg}) into (\ref{Laplf}), we get a closed equation for the
complex function $\tilde{f}:$ 
\begin{equation}
\left[ \left( p+\gamma _{+}-i\tau _{0}\right) \left( p+\gamma _{3}\right) +%
{\displaystyle{\kappa _{0}^{2} \over 2}}%
\right] \tilde{f}\left( p\right) +i\kappa _{0}-%
{\displaystyle{\kappa _{0}^{2} \over 2}}%
\tilde{f}^{\ast }\left( p\right) +\gamma _{-}\left( p+\gamma _{3}\right)
e^{-2i\alpha _{0}}\tilde{f}^{\ast }\left( p+2i\dot{\alpha}_{0}\right) =0
\label{decf}
\end{equation}

Note that the persistence length is determined by $\tilde{g}_{3}\left(
0\right) $ which can be expressed through $\tilde{f}\left( 0\right) -\tilde{f%
}^{\ast }\left( 0\right) $, Eq. (\ref{Laplg}). The latter functions can be
calculated from Eq. (\ref{decf}), which upon substituting $p=-2in\dot{\alpha}%
_{0}$ ($n$ integer), is recast in the standard form of difference equations, 
\begin{equation}
a_{n}\kappa _{0}\tilde{f}\left( -2in\dot{\alpha}_{0}\right) +2i-\kappa _{0}%
\tilde{f}^{\ast }\left( -2in\dot{\alpha}_{0}\right) +2\gamma
_{-}b_{n}e^{-2i\alpha _{0}}\tilde{f}^{\ast }\left[ -2i\left( n-1\right) \dot{%
\alpha}_{0}\right] =0  \label{unkn}
\end{equation}
where we defined 
\begin{equation}
a_{n}=1+2\left[ \gamma _{+}-i\left( \tau _{0}+2n\dot{\alpha}_{0}\right) %
\right] \left( \gamma _{3}-2in\dot{\alpha}_{0}\right) /\kappa
_{0}^{2},\qquad b_{n}=\left( \gamma _{3}-2in\dot{\alpha}_{0}\right) /\kappa
_{0}  \label{anbn}
\end{equation}
Since the persistence length is defined as the average of $\tilde{g}%
_{3}\left( 0\right) $ with respect to $\alpha _{0}$, it is convenient to
introduce dimensionless functions $h_{n}$ as: 
\begin{equation}
h_{n}=\kappa _{0}\int_{0}^{2\pi }\frac{d\alpha _{0}}{2\pi }e^{2in\alpha _{0}}%
\tilde{f}\left( -2in\dot{\alpha}_{0}\right)  \label{hn}
\end{equation}
We multiply Eq. (\ref{unkn}) by $\exp \left( 2in\alpha _{0}\right) $ and
average it with respect to $\alpha _{0}$. Defining the parameter $%
\varepsilon =2\gamma _{-}/\kappa _{0}$ we rewrite Eq. (\ref{unkn}) in the
form 
\begin{equation}
a_{n}h_{n}+2i\delta _{n0}-h_{-n}^{\ast }+\varepsilon b_{n}h_{1-n}^{\ast }=0
\label{anh}
\end{equation}
in which both $h_{n}$ and $h_{m}^{\ast }$ enter. In order to derive closed
equations for the set of $\left\{ h_{n}\right\} $ only, we apply complex
conjugation to the above equation and change $n\longrightarrow -n$. This
yields 
\begin{equation}
a_{-n}^{\ast }h_{-n}^{\ast }-2i\delta _{n0}-h_{n}+\varepsilon b_{n}h_{n+1}=0
\label{anhconj}
\end{equation}
Substituting the equations for $h_{-n}^{\ast }$ and $h_{1-n}^{\ast }$ into (%
\ref{anh}) we find 
\begin{equation}
\begin{array}{c}
\left( a_{n}-1/a_{-n}^{\ast }-\varepsilon ^{2}b_{n}b_{n-1}/a_{1-n}^{\ast
}\right) h_{n}+2i\left( 1-1/a_{-n}^{\ast }\right) \delta _{n0}+ \\ 
2i\varepsilon \delta _{n1}b_{n}/a_{1-n}^{\ast }+\varepsilon
h_{n+1}b_{n}/a_{-n}^{\ast }+\varepsilon h_{n-1}b_{n}/a_{1-n}^{\ast }=0
\end{array}
\label{ahb}
\end{equation}

Let us first consider the case $n\neq 0,1$. Introducing new variables $y_{n}$
by the equality $h_{n+1}=\varepsilon y_{n}h_{n}$ we find 
\begin{equation}
a_{n}-1/a_{-n}^{\ast }-\varepsilon ^{2}b_{n}b_{n-1}/a_{1-n}^{\ast
}+\varepsilon ^{2}y_{n}b_{n}/a_{-n}^{\ast }+y_{n-1}^{-1}b_{n}/a_{1-n}^{\ast
}=0  \label{yn}
\end{equation}
We now define 
\begin{equation}
A_{n}=\left( a_{n}-1/a_{-n}^{\ast }\right) a_{1-n}^{\ast }/b_{n}-\varepsilon
^{2}b_{n-1},\qquad B_{n}=a_{1-n}^{\ast }/a_{-n}^{\ast }  \label{AnBn}
\end{equation}
and get the following recurrence relation, valid for $n=2,3,$ ... 
\begin{equation}
A_{n}+1/y_{n-1}+\varepsilon ^{2}B_{n}y_{n}=0
\end{equation}
We now take $n=2$ in the above equation, and solve for $y_{1}\ $in terms of $%
y_{2}.$ Repeating this procedure (expressing $y_{2}$ in terms of $y_{3}$,
etc.) we can write the solution as a continued fraction 
\begin{equation}
y_{1}=-1/\left( A_{2}-\varepsilon ^{2}B_{2}/\left( A_{3}-\varepsilon
^{2}B_{3}/\left( A_{4}-\cdots \right) \right) \right)  \label{y1}
\end{equation}

Now consider the case $n=1$ in Eq. (\ref{ahb}). Using the definitions of $%
A_{1}$ and $B_{1}$, Eq. (\ref{AnBn}), it can be recast into the form: 
\begin{equation}
\left( A_{1}+\varepsilon ^{2}B_{1}y_{1}\right) h_{1}+2i\varepsilon
+\varepsilon h_{0}=0
\end{equation}
In order to obtain a closed equation for $h_{0}$, we return to Eq. (\ref
{anhconj}) with $n=0,$%
\begin{equation}
a_{0}^{\ast }h_{0}^{\ast }-2i-h_{0}+\varepsilon b_{0}h_{1}=0
\end{equation}
Eliminating $h_{1}$ from the above two equations we find 
\begin{equation}
h_{0}=-2i+\Xi h_{0}^{\ast }  \label{h00}
\end{equation}
where, using Eq. (\ref{y1}), $\Xi $ can be represented as a continued
fraction: 
\begin{eqnarray}
\Xi &=&a_{0}^{\ast }/\left( 1+\varepsilon ^{2}b_{0}/\left( A_{1}+\varepsilon
^{2}B_{1}y_{1}\right) \right) \\
&=&a_{0}^{\ast }/\left( 1+\varepsilon ^{2}b_{0}/\left( A_{1}-\varepsilon
^{2}B_{1}/\left( A_{2}-\varepsilon ^{2}B_{2}/\left( A_{3}-\cdots \right)
\right) \right) \right)
\end{eqnarray}

The solution of Eq. (\ref{h00}) is: 
\begin{equation}
h_{0}=-2i%
{\displaystyle{1-\Xi  \over 1-\left| \Xi \right| ^{2}}}%
\end{equation}
Recall that $h_{0}$ was defined as the integral over $\alpha _{0}$ of the
function $\tilde{f}\left( 0\right) $ (Eq. (\ref{hn})) which, in turn,
determines the Laplace transform at $p=0$ of the correlator $\tilde{g}_{3}$
that appears in the definition of the persistence length, Eq. (\ref{persist}%
). Collecting the above expressions we find: 
\begin{equation}
l_{p}=\int_{0}^{2\pi }\frac{d\alpha _{0}}{\pi }\tilde{g}_{3}\left( 0\right) =%
\frac{2}{\gamma _{3}}\left[ 1-i%
{\displaystyle{1 \over 2}}%
\left( h_{0}-h_{0}^{\ast }\right) \right] =%
{\displaystyle{2\gamma _{3}^{-1} \over 1+\left( \Xi -1\right) ^{-1}+\left( \Xi ^{\ast }-1\right) ^{-1}}}%
\end{equation}

\bigskip

{\Large Acknowledgment }

We would like to thank A. Drozdov for illuminating discussions and D.
Kessler for helpful comments on the manuscript. YR acknowledges support by a
grant from the Israel Science Foundation. SP thanks M. Elbaum for
hospitality during his stay at the Weizmann Institute.

\newpage

{\LARGE {\bf Figure captions}}\bigskip

Figure 1: Schematic drawing of a twisted ribbon--like filament. The vectors
of the physical (${\bf t}_{1},{\bf t}_{2}$) and the Frenet (${\bf b},{\bf n}$%
) triad can be brought into coincidence through rotation by angle $\alpha $,
about the common tangent (${\bf t}_{3}$)$.$

Figure 2: Schematic plot of section of a ribbon--like helix. The
helix--fixed coordinate system ${\bf t}$ at contour point $s^{\prime }$ is
shown. The solid line describes the associated ``rod--like chain'' to which
the coordinate system ${\bf e}$ is attached at point $\sigma $ on its
contour. The points $\sigma $ and $\sigma ^{\prime }$ on the rod--like chain
are the projections of the points $s$ and $s^{\prime }$ respectively.

Figure 3: Three--dimensional plot of the persistence length $l^{\ast }$ as a
function of the dimensionless rate of twist $w$ and of the bare persistence
length $a_{1}$ (logarithmic scale), for a helical filament with spontaneous
curvature $\kappa _{0}=1$, and torsion $\tau _{0}=0.01$ (in arbitrary
units). The bare persistence lengths are $a_{2}=1000$, and $a_{3}=5000$.

Figure 4: Plot of the persistence length $l^{\ast }$ as a function of the
dimensionless rate of twist $w$ for a helical filament with spontaneous
curvature $\kappa _{0}=1$, and torsion $\tau _{0}=0.01$ (in arbitrary
units). The different curves correspond to different bare persistence
lengths: (1) $a_{1}=100$, $a_{2}=a_{3}=5000$, (2) $a_{1}=1$, $%
a_{2}=a_{3}=100 $, (3) $a_{1}=0.1$, $a_{2}=a_{3}=10$, (4) $a_{1}=0.01$, $%
a_{2}=a_{3}=10$. A magnified view of the region of small twist rates is
shown in the insert.

Figure 5: Three--dimensional plot of the persistence length $l^{\ast }$ as a
function of the dimensionless rate of twist $w$ and of the spontaneous
curvature $\kappa _{0}$, for a helical filament with spontaneous torsion $%
\tau _{0}=1$ (in arbitrary units). The bare persistence lengths are $%
a_{1}=500$, $a_{2}=1$ and $a_{3}=500$.


\begin{references}
\bibitem{comm}  Strictly speaking, this derivation applies only to straight
rods but the results are applicable to our case as long as the radii of
curvature associated with the deformations are much larger than the
thickness of the rod.
\end{references}
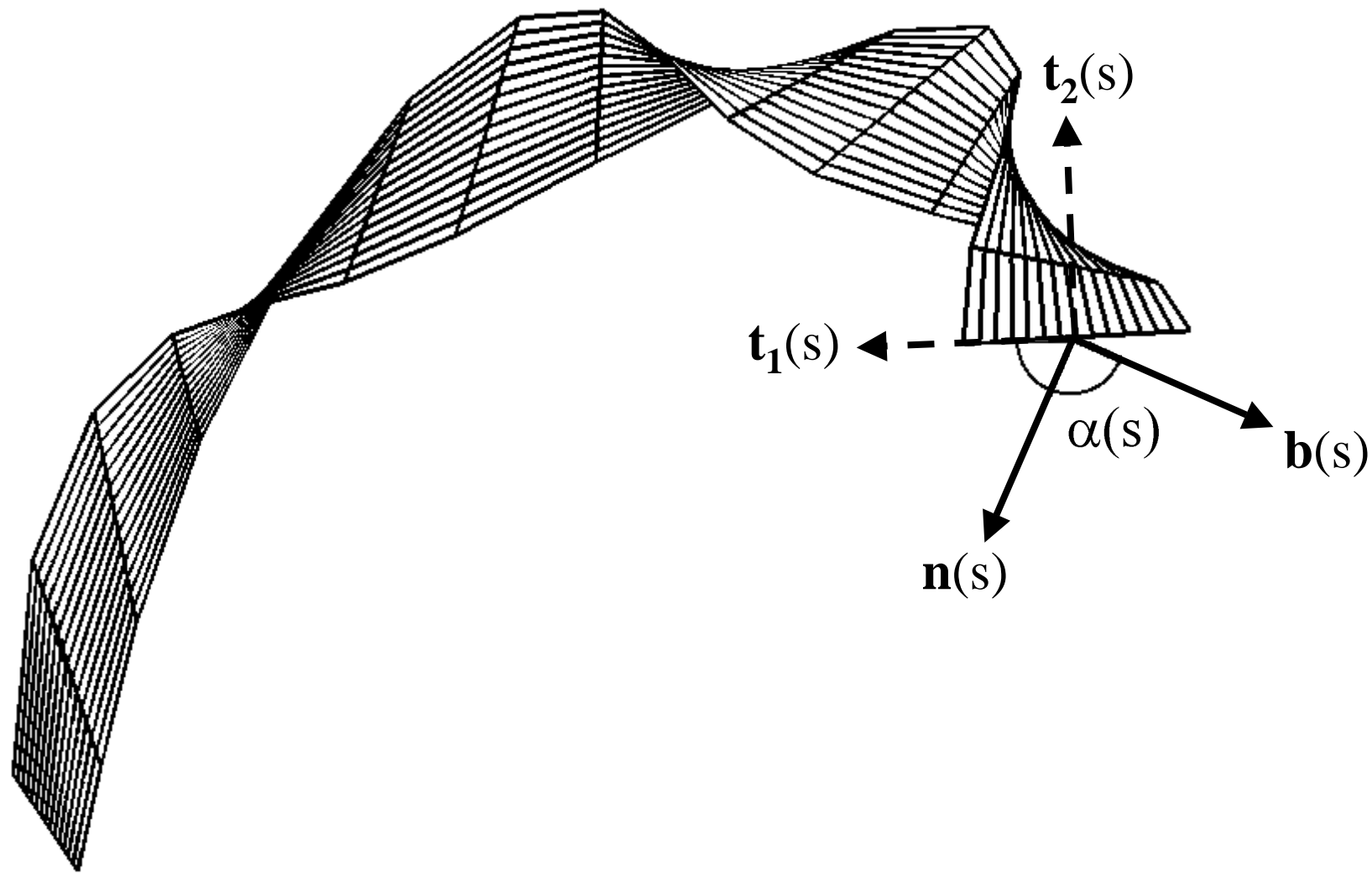
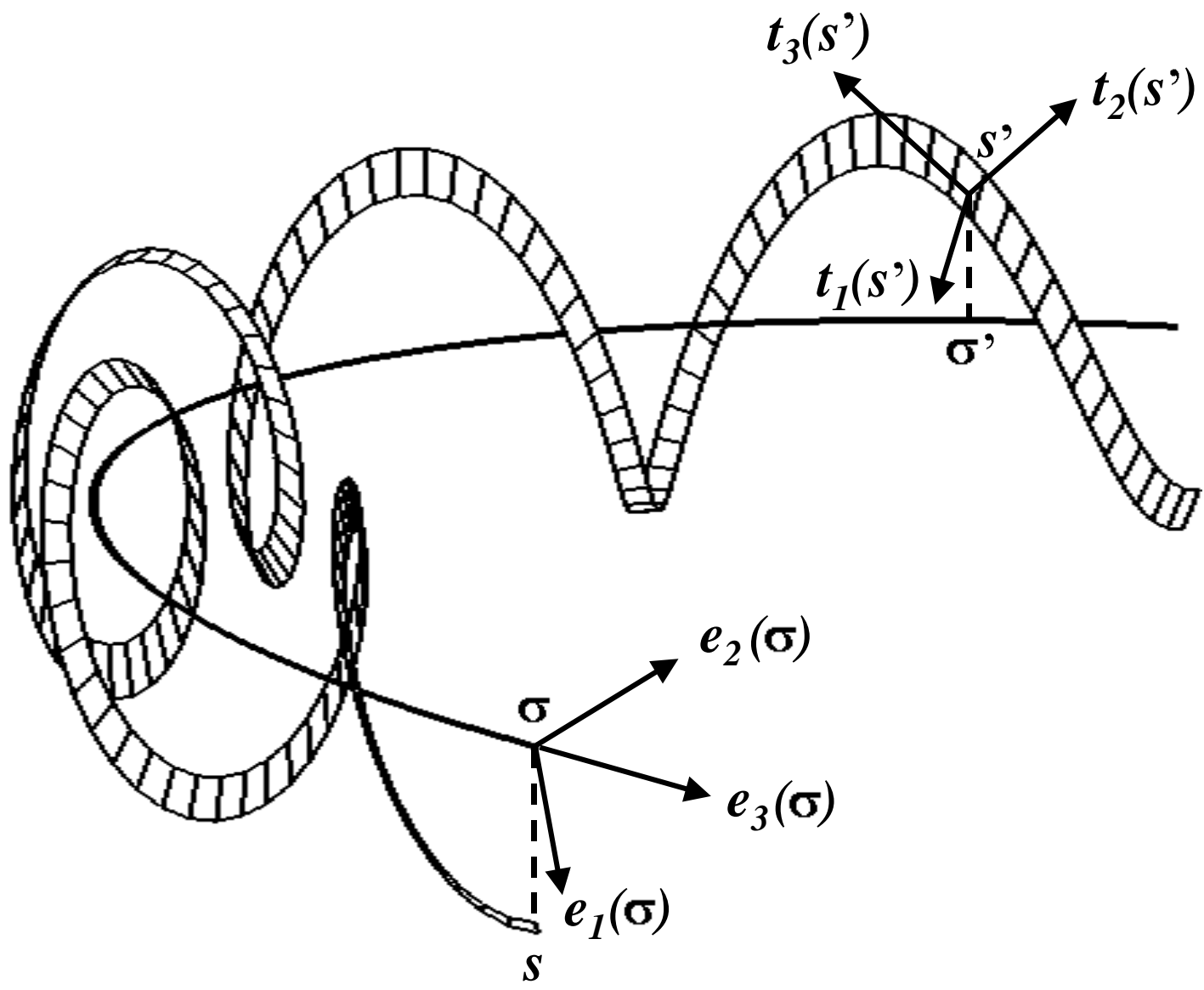
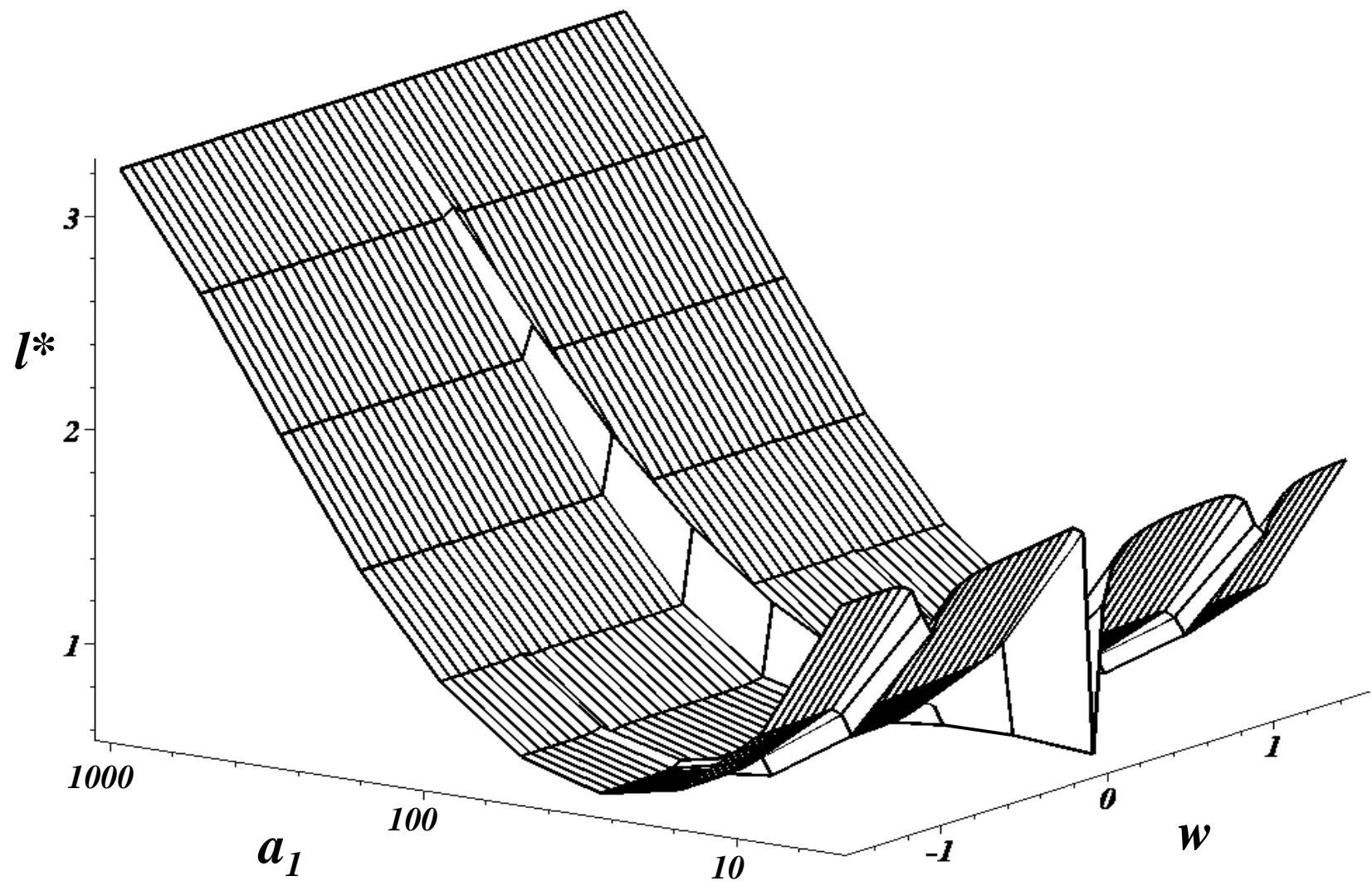
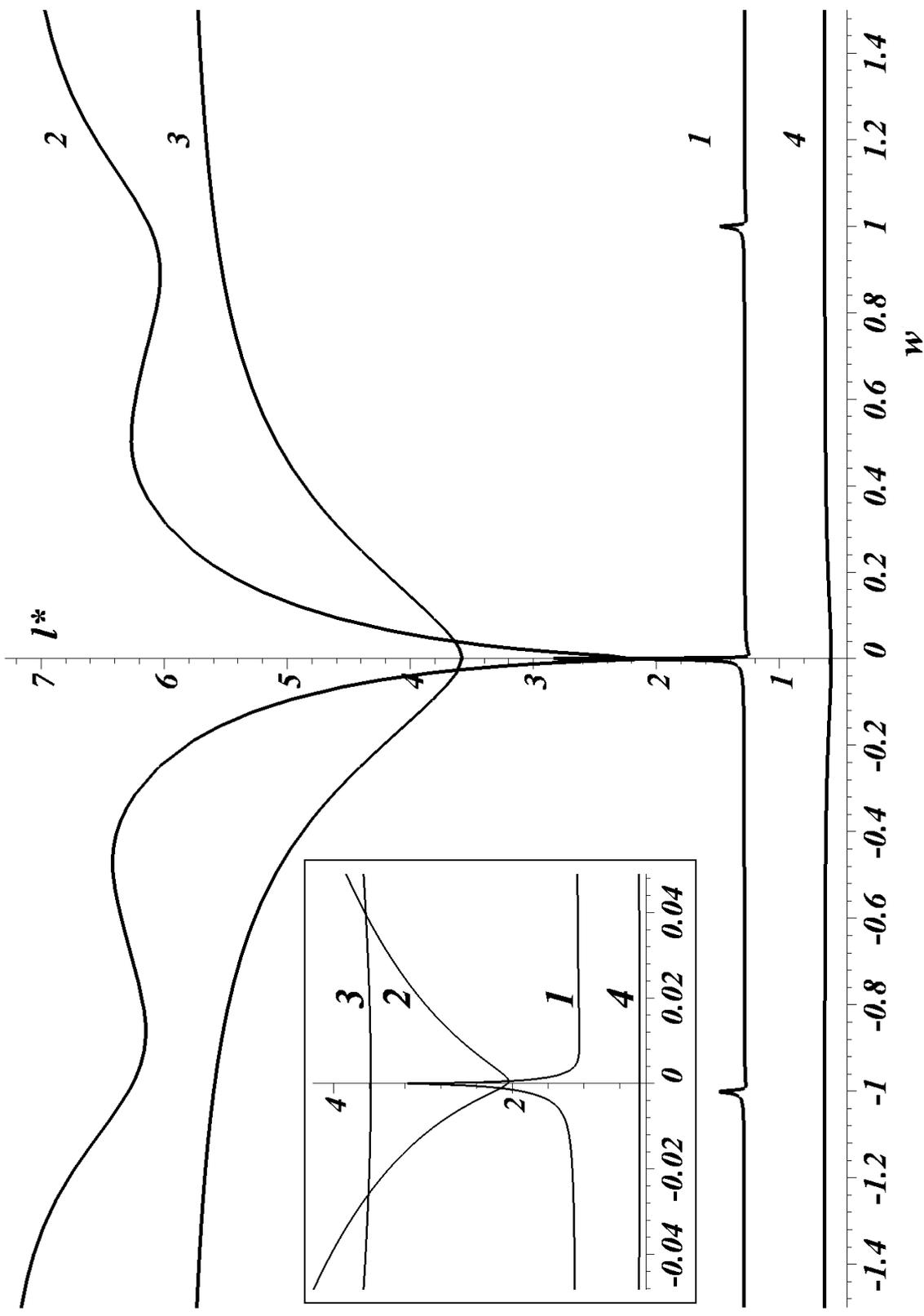
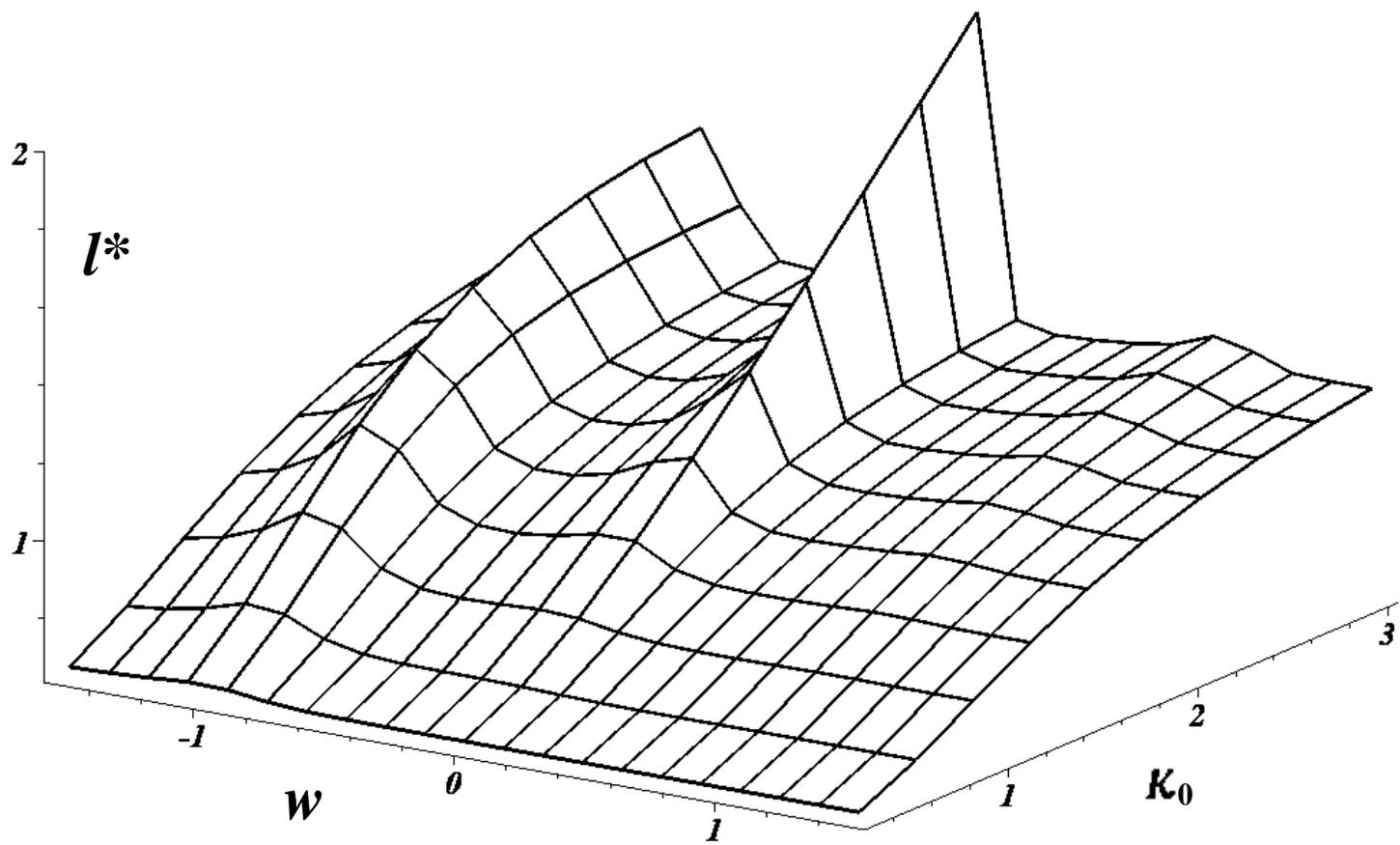
\end{document}